\documentclass[journal, twoside]{IEEEtran}

\usepackage[group-separator={,}]{siunitx}
\usepackage{graphicx}
\usepackage{amsmath}
\usepackage{xcolor}

\usepackage[acronym,nomain]{glossaries}
\glsdisablehyper
\newcommand{\SetCapsType}{normalcaps}

\makeatletter
\providecommand{\SetCapsType}{smallcaps}

\long\def\@scTrue{smallcaps}
\long\def\@scFalse{normalcaps}
\newcommand{\acroSCaps}[1]{%
 \begingroup
  \ifx\SetCapsType\@scTrue 
    \textsc{#1}%
  \else
    \MakeUppercase{#1}%
  \fi
  \endgroup
}
\makeatother

\newcommand{\nAcronym}[4][]{%
	\newacronym[#1]{#2}{#3}{#4}
}

\makeatletter
\@ifpackageloaded{babel}{%
    \newcommand{\usuk}[2]{%
        \iflanguage{USenglish}{#1}{#2}%
    }%
}{%
    \newcommand{\usuk}[2]{%
        #1%
    }%
}%
\makeatother

\usepackage[shortcuts]{extdash}
\newcommand{\qam}[1]{
    \ifglsused{QAM}%
        {#1\=/\gls{QAM}}%
        {#1\=/ary \gls{QAM}%
    }%
}%

\nAcronym{2A8PSK}{\acroSCaps{2a8psk}}{2-ary amplitude 8-ary phaseshift keying}

\nAcronym{3CCMCF}{\acroSCaps{3cc-mcf}}{3-core coupled-core multi-core fiber}

\nAcronym{4D}{\acroSCaps{4d}}{four-dimensional}
\nAcronym{4D64PRS}{\acroSCaps{4d-64prs}}{\usuk{four-dimensional 64-ary polarization-ring-switching}{four-dimensional 64-ary polarisation-ring-switching}}

\nAcronym{5B4D2A8PSK}{\acroSCaps{5b4d-2a8psk}}{5-bit four-dimensional two-amplitude 8-ary phase-shift keying}

\nAcronym{6B4D2A8PSK}{\acroSCaps{6b4d-2a8psk}}{6-bit four-dimensional two-amplitude 8-ary phase-shift keying}

\nAcronym{8D}{\acroSCaps{8d}}{eight-dimensional}\nAcronym{8D2048PRS}{\acroSCaps{8d-2048prs}}{eight-dimensional 2048-ary polarization-ring-switching}
\nAcronym{8D2048PRST1}{\acroSCaps{8d-2048prs-t1}}{eight-dimensional 2048-ary polarization-ring-switching type 1}
\nAcronym{8D2048PRST2}{\acroSCaps{8d-2048prs-t2}}{eight-dimensional 2048-ary polarization-ring-switching type 2}

\nAcronym{ABC}{\acroSCaps{abc}}{automatic bias controller}
\nAcronym{ADC}{\acroSCaps{adc}}{analog-to-digital converter}
\nAcronym{AIR}{\acroSCaps{air}}{achievable information rate}
\nAcronym{AOM}{\acroSCaps{aom}}{acoustic optical modulator}
\nAcronym{API}{\acroSCaps{api}}{application programming interface}
\nAcronym{AR}{\acroSCaps{ar}}{achievable rate}
\nAcronym{ASE}{\acroSCaps{ase}}{amplified spontaneous emission}
\nAcronym{ASK}{\acroSCaps{ask}}{amplitude-shift keying}
\nAcronym{ASIC}{\acroSCaps{asic}}{application-specific integrated circuit}
\nAcronym{ARRWG}{\acroSCaps{a}rr\acroSCaps{wg}}{arrayed-waveguide grating}
\nAcronym{AWG}{\acroSCaps{awg}}{arbitrary-waveform generator}
\nAcronym{AWGN}{\acroSCaps{awgn}}{additive white Gaussian noise}

\nAcronym{BCH}{\acroSCaps{bch}}{Bose-Chaudhuri-Hocquenghem}
\nAcronym{BER}{\acroSCaps{ber}}{bit error rate}
\nAcronym{BICM}{\acroSCaps{bicm}}{bit-interleaved coded modulation}
\nAcronym{BMD}{\acroSCaps{bmd}}{bit-metric decoding}
\nAcronym{BPD}{\acroSCaps{bpd}}{balanced photo-diode}
\nAcronym{BPF}{\acroSCaps{bpf}}{bandpass filter}
\nAcronym{BPS}{\acroSCaps{bps}}{blind phase search}
\nAcronym{BPSK}{\acroSCaps{bpsk}}{binary phase-shift keying}
\nAcronym{BRGC}{\acroSCaps{brgc}}{binary reflected Gray code}
\nAcronym{BTB}{\acroSCaps{btb}}{back-to-back}

\nAcronym{CAGR}{\acroSCaps{cagr}}{compound annual growth rate}
\nAcronym{CCDM}{\acroSCaps{ccdm}}{constant composition distribution matching}
\nAcronym{CCF}{\acroSCaps{ccf}}{\usuk{coupled-core fiber}{coupled-core fibre}}
\nAcronym{CD}{\acroSCaps{cd}}{chromatic dispersion}
\nAcronym{CIR}{\acroSCaps{cir}}{channel impulse response}
\nAcronym{CMA}{\acroSCaps{cma}}{constant modulus algorithm}
\nAcronym{CMUX}{\acroSCaps{cmux}}{core multiplexer}
\nAcronym{ChUT}{\acroSCaps{chut}}{channel under test}
\nAcronym{CUT}{\acroSCaps{cut}}{channel under test}
\nAcronym{CPE}{\acroSCaps{cpe}}{carrier phase estimation}
\nAcronym{CPU}{\acroSCaps{cpu}}{central processing unit}
\nAcronym{CSPR}{\acroSCaps{cspr}}{carrier-to-signal power ratio}
\nAcronym{CW}{\acroSCaps{cw}}{continuous wave}

\nAcronym{DA}{\acroSCaps{da}}{driver amplifier}
\nAcronym{DAC}{\acroSCaps{dac}}{digital-to-analog converter}
\nAcronym{DCF}{\acroSCaps{dcf}}{\usuk{dispersion compensated fiber}{dispersion compensated fibre}}
\nAcronym{DCI}{\acroSCaps{dci}}{data center interconnect}
\nAcronym{DDLMS}{\acroSCaps{dd-lms}}{decision-directed least mean square}
\nAcronym{DFB}{\acroSCaps{dfb}}{distributed feedback}
\nAcronym{DGD}{\acroSCaps{dgd}}{differential group delay}
\nAcronym{DH}{\acroSCaps{dh}}{digital holography}
\nAcronym{DM}{\acroSCaps{dm}}{distribution matcher}
\nAcronym{DMA}{\acroSCaps{dma}}{direct memory access}
\nAcronym{DMD}{\acroSCaps{dmd}}{differential mode delay}
\nAcronym{DMG}{\acroSCaps{dmg}}{differential modal gain}
\nAcronym{DMGD}{\acroSCaps{dmgd}}{differential mode group delay}
\nAcronym{DML}{\acroSCaps{dml}}{directly-modulated laser}
\nAcronym{DP}{\acroSCaps{dp}}{\usuk{dual-polarization}{dual-polarisation}}
\nAcronym{DPC}{\acroSCaps{dpc}}{digital pre-compensation}
\nAcronym{DPE}{\acroSCaps{dpe}}{digital pre-emphasis}
\nAcronym{DPIQ}{\acroSCaps{dp-iqm}}{\usuk{dual-polarization IQ-modulator}{dual-polarisation IQ-modulator}}
\nAcronym{DPLL}{\acroSCaps{dpll}}{digital phase-locked loop}
\nAcronym{DQPSK}{\acroSCaps{dqpsk}}{differential quaternary phase-shift-keying}
\nAcronym{DRA}{\acroSCaps{dra}}{distributed Raman amplifier}
\nAcronym{DRE}{\acroSCaps{dre}}{digital resolution enhancer}
\nAcronym{DSF}{\acroSCaps{dsf}}{\usuk{dispersion-shifted fiber}{dispersion-shifted fibre}}
\nAcronym{DSO}{\acroSCaps{dso}}{digital sampling oscilloscope}
\nAcronym{DSP}{\acroSCaps{dsp}}{digital signal processing}
\nAcronym{DUT}{\acroSCaps{dut}}{device-under-test}
\nAcronym{DWDM}{\acroSCaps{dwdm}}{dense wavelength-division multiplexing}

\nAcronym{EAM}{\acroSCaps{eam}}{electro-absorption modulator}
\nAcronym{ECL}{\acroSCaps{ecl}}{external cavity laser}
\nAcronym{ED}{\acroSCaps{ed}}{Eucledian distance}
\nAcronym{EDFA}{\acroSCaps{edfa}}{\usuk{erbium-doped fiber amplifier}{erbium-doped fibre amplifier}}
\nAcronym{ENOB}{\acroSCaps{enob}}{effective number of bits}
\nAcronym{ESS}{\acroSCaps{ess}}{enumerative sphere shaping}

\nAcronym{FD}{\acroSCaps{fd}}{frequency domain}
\nAcronym{FDE}{\acroSCaps{fde}}{\usuk{frequency domain equalizer}{frequency domain equaliser}}
\nAcronym{FEC}{\acroSCaps{fec}}{forward error correction}
\nAcronym{FFT}{\acroSCaps{fft}}{fast Fourier transform}
\nAcronym{FIR}{\acroSCaps{fir}}{finite impulse response}
\nAcronym{FMF}{\acroSCaps{fmf}}{\usuk{few-mode fiber}{few-mode fibre}}
\nAcronym[plural=FM-MCF, firstplural=\usuk{few-mode multi-core fibers}{few-mode multi-core fibres}]{FM-MCF}{\acroSCaps{fm-mcf}}{\usuk{few-mode multi-core fiber}{few-mode multi-core fibre}}
\nAcronym{FPGA}{\acroSCaps{fpga}}{field-programmable gate array}
\nAcronym{FWM}{\acroSCaps{fwm}}{four-wave mixing}

\nAcronym{GD}{\acroSCaps{gd}}{group delay}
\nAcronym{GI}{\acroSCaps{gi}}{graded-index}
\nAcronym{GFF}{\acroSCaps{gff}}{gain flattening filter}
\nAcronym{GIMMF}{\acroSCaps{gi-mmf}}{\usuk{graded-index multi-mode fiber}{graded-index multi-mode fibre}}
\nAcronym{GMI}{\acroSCaps{gmi}}{\usuk{generalized mutual information}{generalised mutual information}}
\nAcronym{GNSE}{\acroSCaps{gnse}}{generalized nonlinear Schr\"{o}dinger equation}
\nAcronym{GPU}{\acroSCaps{gpu}}{graphics processing unit}
\nAcronym{GS}{\acroSCaps{gs}}{geometric shaping}
\nAcronym{GV}{\acroSCaps{gv}}{group-velocity}
\nAcronym{GVD}{\acroSCaps{gvd}}{group-velocity dispersion}
\nAcronym{GPIO}{\acroSCaps{gpio}}{general purpose input output}

\nAcronym{HDFEC}{\acroSCaps{hd-fec}}{hard decision forward error correction}

\nAcronym[plural=IL, firstplural=insertion losses (\textsc{il})]{IL}{\acroSCaps{il}}{insertion loss}
\nAcronym{IFFT}{\acroSCaps{ifft}}{inverse fast Fourier transform}
\nAcronym{IMDD}{\acroSCaps{im/dd}}{intensity-modulation direct-detection}
\nAcronym{IQM}{\acroSCaps{iqm}}{in-phase and quadrature modulator}
\nAcronym{ISI}{\acroSCaps{isi}}{intersymbol interference}

\nAcronym{KK}{\acroSCaps{kk}}{Kramers-Kronig}

\nAcronym{LCOS}{\acroSCaps{LCoS}}{liquid crystal on silicon}
\nAcronym{LDPC}{\acroSCaps{ldpc}}{low-density parity-check}
\nAcronym{LEAF}{\acroSCaps{leaf}}{\usuk{large effective area fiber}{large effective area fibre}}
\nAcronym{LFSR}{\acroSCaps{lfsr}}{linear-feedback shift register}
\nAcronym{LMS}{\acroSCaps{lms}}{least means square}
\nAcronym{LLR}{\acroSCaps{llr}}{log-likelihood ratio}
\nAcronym{LO}{\acroSCaps{lo}}{local oscillator}
\nAcronym{LP}{\acroSCaps{lp}}{\usuk{linearly polarized}{linearly polarised}}
\nAcronym{LSPS}{\acroSCaps{lsps}}{\usuk{loop-synchronized polarization scrambler}{loop-synchronised polarisation scrambler}}
\nAcronym{LUT}{\acroSCaps{lut}}{lookup table}

\nAcronym{MB}{\acroSCaps{mb}}{Maxwell-Bolzmann}
\nAcronym{MCF}{\acroSCaps{mcf}}{\usuk{multi-core fiber}{multi-core fibre}}
\nAcronym{MDG}{\acroSCaps{mdg}}{mode dependent gain}
\nAcronym[plural=MDL, firstplural=mode-dependent losses (\textsc{mdl})]{MDL}{\acroSCaps{mdl}}{mode-dependent loss}
\nAcronym{MDM}{\acroSCaps{mdm}}{mode division multiplexing}
\nAcronym{MEMS}{\acroSCaps{mems}}{micro-electro-mechanical systems}
\nAcronym{MF}{\acroSCaps{mf}}{matched filter}
\nAcronym{MI}{\acroSCaps{mi}}{mutual information}
\nAcronym{MIMO}{\acroSCaps{mimo}}{multiple-input multiple-output}
\nAcronym{ML}{\acroSCaps{ml}}{machine learning}
\nAcronym{MMA}{\acroSCaps{mma}}{multi-modulus algorithm}
\nAcronym{MMF}{\acroSCaps{mmf}}{\usuk{multi-mode fiber}{multi-mode fibre}}
\nAcronym{MMSE}{\acroSCaps{mmse}}{minimum mean squared error}
\nAcronym{MP}{\acroSCaps{mp}}{minimum phase}
\nAcronym{MPLC}{\acroSCaps{mplc}}{multi-plane light converter}
\nAcronym{MSE}{\acroSCaps{mse}}{mean squared error}
\nAcronym{MUX}{\acroSCaps{mux}}{multiplexer}
\nAcronym{MZM}{\acroSCaps{mzm}}{Mach-Zehnder modulator}
\nAcronym{MZI}{\acroSCaps{mzi}}{Mach-Zehnder interferometer}

\nAcronym{NA}{\acroSCaps{na}}{numerical aperture}
\nAcronym{NF}{\acroSCaps{nf}}{noise figure}
\nAcronym{NGMI}{\acroSCaps{ngmi}}{\usuk{normalized generalized mutual information}{normalised generalised mutual information}}
\nAcronym{NLSE}{\acroSCaps{nlse}}{nonlinear Schr\"{o}ding equation}
\nAcronym{NIR}{\acroSCaps{nir}}{near-infrared}

\nAcronym{OBTB}{\acroSCaps{obtb}}{optical back-to-back}
\nAcronym{ODE}{\acroSCaps{ode}}{ordinary differential equation}
\nAcronym{ODL}{\acroSCaps{odl}}{optical delay line}
\nAcronym{OFC}{\acroSCaps{ofc}}{Optical Fiber Communications Conference}
\nAcronym{OFDR}{\acroSCaps{ofdr}}{optical frequency-domain reflectometer}
\nAcronym{OFDM}{\acroSCaps{ofdm}}{orthogonal frequency division multiplexing}
\nAcronym{OH}{\acroSCaps{oh}}{overhead}
\nAcronym{OMFT}{\acroSCaps{omft}}{optical multi-format transmitter}
\nAcronym{OOK}{\acroSCaps{ook}}{on-off keying}
\nAcronym{OPLL}{\acroSCaps{opll}}{optical phase-locked loop}
\nAcronym{OSA}{\acroSCaps{osa}}{\usuk{optical spectrum analyzer}{optical spectrum analyser}}
\nAcronym{OSNR}{\acroSCaps{osnr}}{optical signal-to-noise ratio}
\nAcronym{OTDR}{\acroSCaps{otdr}}{optical time-domain reflectometer}
\nAcronym{OTF}{\acroSCaps{otf}}{optical tunable filter}
\nAcronym{OVNA}{\acroSCaps{ovna}}{\usuk{optical vector network analyzer}{optical vector network analyser}}

\nAcronym{PAM}{\acroSCaps{pam}}{pulse-amplitude modulation}
\nAcronym{PAS}{\acroSCaps{pas}}{probabilistic amplitude shaping}
\nAcronym{PAPR}{\acroSCaps{papr}}{peak-to-avarage power ratio}
\nAcronym{PBC}{\acroSCaps{pbc}}{\usuk{polarization beam combiner}{polarisation beam combiner}}
\nAcronym{PBS}{\acroSCaps{pbs}}{polarization beam splitter}
\nAcronym{PCVD}{\acroSCaps{pcvd}}{plasma chemical vapor depostion}
\nAcronym{PD}{\acroSCaps{pd}}{photodiode}
\nAcronym{PDL}{\acroSCaps{pdl}}{polarization-dependent loss}
\nAcronym{PDM}{\acroSCaps{pdm}}{\usuk{polarization-division multiplexing}{polarisation-division multiplexing}}
\nAcronym{PL}{\acroSCaps{pl}}{photonic lantern}
\nAcronym{PMBPSK}{\acroSCaps{pm-bpsk}}{polarization-multiplexed binary phase-shift-keying}
\nAcronym{PMQPSK}{\acroSCaps{pm-qpsk}}{polarization-multiplexed quaternary phase-shift-keying}
\nAcronym{PM8QAM}{\acroSCaps{pm-8qam}}{polarization-multiplexed 8-ary quadrature amplitude modulation}
\nAcronym{PMD}{\acroSCaps{pmd}}{polarization mode dispersion}
\nAcronym{PNOB}{\acroSCaps{pnob}}{physical number of bits}
\nAcronym{PON}{\acroSCaps{pon}}{passive-optical network}
\nAcronym{PRBS}{\acroSCaps{prbs}}{pseudorandom bit sequence}
\nAcronym{PS}{\acroSCaps{ps}}{probabilistic shaping}
\nAcronym{PSK}{\acroSCaps{psk}}{phase-shift keying}
\nAcronym{PSP}{\acroSCaps{psp}}{\usuk{principal states of polarization}{principal states of polarisation}}

\nAcronym{QAM}{\acroSCaps{qam}}{quadrature amplitude modulation}
\nAcronym{QPSK}{\acroSCaps{qpsk}}{quaternary phase-shift-keying}

\nAcronym{RAM}{\acroSCaps{ram}}{random-access memory}
\nAcronym{RF}{\acroSCaps{rf}}{radio frequency}
\nAcronym{RRC}{\acroSCaps{rrc}}{root-raised-cosine}
\nAcronym{ROADM}{\acroSCaps{roadm}}{reconfigurable optical add-drop multiplexer}
\nAcronym{ROI}{\acroSCaps{roi}}{region of interest}

\nAcronym{SA}{\acroSCaps{sa}}{simulated annealing}
\nAcronym{SamPerSym}{\acroSCaps{sps}}{samples per symbol}
\nAcronym{SBS}{\acroSCaps{sbs}}{stimulated Brillouin scattering}
\nAcronym{SDFEC}{\acroSCaps{sd-fec}}{soft decision forward error correction}
\nAcronym{SDM}{\acroSCaps{sdm}}{space-division multiplexing}
\nAcronym{SE}{\acroSCaps{se}}{spectral efficiency}
\nAcronym{SER}{\acroSCaps{ser}}{symbol error rate}
\nAcronym{SI}{\acroSCaps{si}}{step index}
\nAcronym{SLM}{\acroSCaps{slm}}{spatial light modulator}
\nAcronym{SMF}{\acroSCaps{smf}}{\usuk{single-mode fiber}{single-mode fibre}}
\nAcronym{SNR}{\acroSCaps{snr}}{signal-to-noise ratio}
\nAcronym{SOA}{\acroSCaps{soa}}{semiconductor optical amplifier}
\nAcronym[firstplural=\usuk{states of polarization (\acroSCaps{sop})}{states of polarisation (\acroSCaps{sop})}]{SOP}{\acroSCaps{sop}}{\usuk{state of polarization}{state of polarisation}}
\nAcronym{SPM}{\acroSCaps{spm}}{self-phase modulation}
\nAcronym{SPS}{\acroSCaps{sps}}{\usuk{synchronized polarization scrambler}{synchronised polarisation scrambler}}
\nAcronym{SRS}{\acroSCaps{srs}}{stimulated Raman scattering}
\nAcronym{SSBI}{\acroSCaps{ssbi}}{signal-signal beat interference}
\nAcronym{SSFM}{\acroSCaps{ssfm}}{split-step Fourier method}
\nAcronym{SSMF}{\acroSCaps{ssmf}}{\usuk{standard single-mode fiber}{standard single-mode fibre}}
\nAcronym{STL}{\acroSCaps{stl}}{swept tunable laser}
\nAcronym{SVD}{\acroSCaps{svd}}{singular value decomposition}
\nAcronym{SWI}{\acroSCaps{swi}}{swept wavelength interferometry}

\nAcronym{TD}{\acroSCaps{td}}{time domain}
\nAcronym{TDE}{\acroSCaps{tde}}{\usuk{time domain equalizer}{time domain equaliser}}
\nAcronym{TDMSDM}{\acroSCaps{tdm-sdm}}{time-domain multiplexed space-division multiplexing}
\nAcronym{TE}{\acroSCaps{te}}{transverse electric}
\nAcronym{TIA}{\acroSCaps{tia}}{trans-impedance amplifier}
\nAcronym{TM}{\acroSCaps{TM}}{transverse magnetic}
\nAcronym{TH4D}{\acroSCaps{th-4d}}{time domain hybrid four-dimensional}
\nAcronym{TH4D2A8PSK}{\acroSCaps{th-4d-2a8psk}}{time domain hybrid four-dimensional two-amplitude eight-phase shift keying}

\nAcronym{VCSEL}{\acroSCaps{vcsel}}{vertical-cavity surface emitting laser}
\nAcronym{VOA}{\acroSCaps{voa}}{variable optical attenuator}

\nAcronym{WDM}{\acroSCaps{wdm}}{wavelength-division multiplexing}
\nAcronym{WGA}{\acroSCaps{wga}}{weakly guiding approximation}
\nAcronym{WGN}{\acroSCaps{wgn}}{white Gaussian noise}
\nAcronym{WSS}{\acroSCaps{wss}}{wavelength selective switch}

\nAcronym{XPM}{\acroSCaps{xpm}}{cross-phase modulation}
\nAcronym{XT}{\acroSCaps{xt}}{cross-talk}

\nAcronym{3DWG}{\acroSCaps{3dwg}}{3D-waveguide}

\usepackage[style=ieee, maxnames=3, defernumbers=true, dashed=false]{biblatex}
\bibliography{ref} 

\DeclareBibliographyCategory{ignore}

\usepackage{lipsum}

\usepackage{caption}
\usepackage{subcaption}

\usepackage[capitalise]{cleveref}

\ifCLASSINFOpdf
\else
\fi

\makeatletter
\let\blx@rerun@biber\relax
\makeatother

\begin{document}
%
\title{Real-time 10,000~km Straight-line Transmission using a Software-defined GPU-Based Receiver}
%
%
%

\author{Sjoerd~van~der~Heide,~\IEEEmembership{Student Member,~IEEE,}
        Ruben~S.~Luis,~\IEEEmembership{Senior Member,~IEEE,}
        Benjamin~J.~Puttnam,~\IEEEmembership{Member,~IEEE,}
        Georg~Rademacher,~\IEEEmembership{Senior Member,~IEEE,}
        Ton~Koonen,~\IEEEmembership{Fellow,~IEEE,}
        Satoshi~Shinada,~\IEEEmembership{Member,~IEEE,}
        Yohinari~Awaji,~\IEEEmembership{Member,~IEEE,}
        Hideaki~Furukawa,~\IEEEmembership{Member,~IEEE,}
        Chigo~Okonkwo,~\IEEEmembership{Senior Member,~IEEE}
\thanks{Sjoerd~van~der~Heide, Ton~Koonen, and Chigo~Okonkwo are with the High-Capacity Optical Transmission Laboratory, Electro-Optical Communications Group, Eindhoven University of Technology, PO Box 513, 5600 MB, Eindhoven, the Netherlands. (e-mail: s.p.v.d.heide@tue.nl, c.m.okonkwo@tue.nl, a.m.j.koonen@tue.nl). Ruben~S.~Luis, Benjamin~J.~Puttnam, Georg~Rademacher, Satoshi~Shinada, Yohinari~Awaji, and Hideaki~Furukawa are with the National Institute of Information and Communications Technology, Photonic Network System Laboratory, 4-2-1, Nukui-Kitamachi, Koganei, Tokyo, 184-8795,  Japan (e-mail: rluis@nict.go.jp).}%
\thanks{Manuscript received xxxx; revised xxxx; accepted xxxx. Date of publication xxxx. This work was partly supported by the Dutch NWO Gravitation Program on Research Center for Integrated Nanophotonics under Grant GA~024.002.033.}
}

%
%

\markboth{Photonics Technology Letters,~Vol.~xx, No.~x, xxxx~2021}%
{TBA}
%

\IEEEpubid{0000--0000/00\$00.00~\copyright~2021 IEEE}


\maketitle

\begin{abstract}
Real-time 10,000~km transmission over a straight-line link is achieved using a software-defined multi-modulation format receiver implemented on a commercial off-the-shelf general-purpose graphics processing unit (GPU). Minimum phase 1~GBaud \qam{4} signals are transmitted over 10,000~km and successfully received after detection with a Kramers-Kronig (KK) coherent receiver. \qam{8-, 16-, 32-, and 64} are successfully transmitted over 7600, 5600, 3600, and 1600~km, respectively.
\end{abstract}


%
\IEEEpeerreviewmaketitle

\section{Introduction}
\label{sec:introduction}

Real-time \gls{DSP} for optical communications has traditionally taken form of \glspl{ASIC} and \glspl{FPGA} \cite{Randel_FPGA_2015,Beppu_FPGA_2020,Beppu_2_FPGA_2020}. \Glspl{GPU} have experienced more than a decade of steady exponential growth in terms of computing capability (45\% year-on-year \cite{winzer_scaling_2017}) and energy efficiency (25\% year-on-year \cite{sun_summarizing_2019}), making them a potential alternative to \glspl{ASIC} and \glspl{FPGA}. These improvements combined with low development effort and rapid turnaround have enabled recent demonstrations of real-time \gls{DSP} for optical communications \cite{Li_realtime_LDPC, suzuki_fec,suzuki_phy_2018, suzuki_phy_2019,suzuki_phy_2020,suzuki_real-time_2020,ECOC_realtimeGPU,gpuJLT,OFC_realtime10k}. These demonstrations include real-time \gls{FEC} decoding \cite{Li_realtime_LDPC, suzuki_fec}, physical-layer functionality \cite{suzuki_phy_2018, suzuki_phy_2019, suzuki_phy_2020}, and \gls{DQPSK} detection \cite{suzuki_real-time_2020}. We introduced a real-time flexible multi-modulation format receiver for directly detected pulse-amplitude modulated signals and \gls{KK} \cite{mecozzi_kramers_2016} coherently detected \gls{MP} \cite{mecozzi_kramers_2016} \gls{QAM} signals and verified its operation over a field-deployed link in \cite{ECOC_realtimeGPU,gpuJLT}. The potential to support long distance links with digital compensation of linear transmission impairments, such as chromatic dispersion, was further demonstrated in \cite{OFC_realtime10k}, where this system was shown to support transmission up to \SI{10000}{km}.

This letter revisits and extends the results presented in \cite{OFC_realtime10k}. A real-time flexible multi-modulation format receiver is experimentally evaluated for long-haul optical communications using a \SI{10000}{km} straight-line link consisting of 100 spans with an average length of \SI{100}{km}. The real-time \gls{DSP} is implemented on a commercial off-the-shelf \gls{GPU} and is able to receive \gls{MP} \qam{N} signals detected with a \gls{KK} coherent receiver. The signal processing chain includes the \gls{KK} algorithm as well as equalization, compensating for chromatic dispersion in frequency domain. Successful transmission of \SI{1}{GBaud} \qam{4-, 8-, 16-, 32-, 64} is shown up to \SIlist{10000;7600;5600;3600;1600}{km}, respectively. In addition to the outcomes presented in \cite{OFC_realtime10k}, we address achievable throughput as the critical parameter of optimization for software-defined multi-modulation format systems. Furthermore, to the authors' knowledge, the use of \gls{MP} signals for transmission over such long distances has never been performed, we address the impact of \gls{CSPR} for this transmission scenario. The results demonstrate the potential for \glspl{GPU} in long-haul optical communication systems, in particular for datacenter-to-datacenter applications without the use of third-party hardware.
\IEEEpubidadjcol
\section{Experimental Setup}
\label{sec:expsetup}

\begin{figure*}
\includegraphics[width=\linewidth]{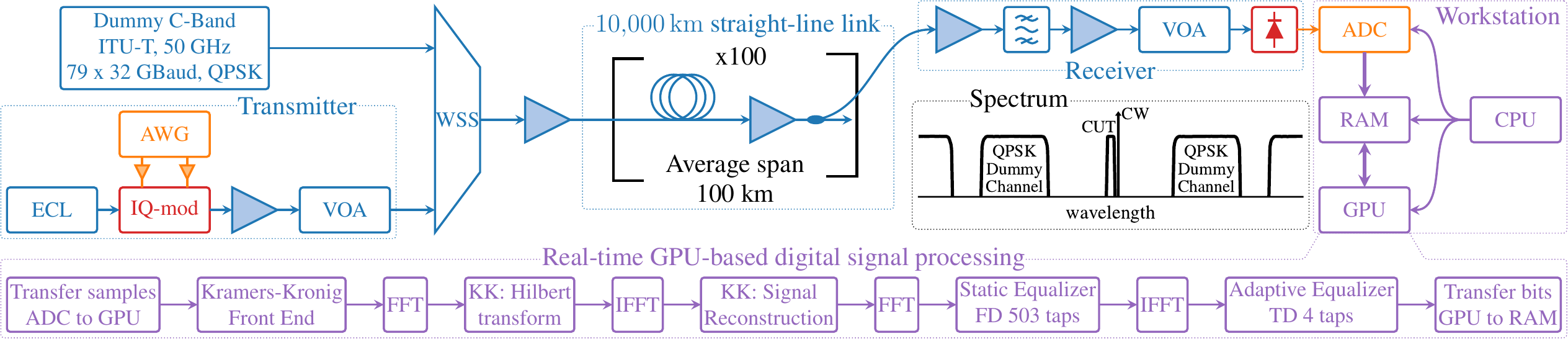}
\caption{Experimental setup including the \SI{10000}{km} straight-line link and a schematic of the main DSP blocks.}
\label{fig:setup} 
\end{figure*}

\cref{fig:setup} shows the experimental setup, including the real-time \gls{DSP} chain. The \gls{MP} \qam{N} test signals are modulated onto the lightwave originating from a \SI{100}{kHz} \gls{ECL} by an \gls{IQM} driven by a 2-channel \SI{12}{GS/s} \gls{AWG}, equipped with RF drivers. The modulator arms are kept at the minimum optical output bias point using bias tees and voltage sources. To ensure that the signal is \gls{MP}, a carrier tone is inserted digitally with  \gls{CSPR} controlled by modifying the power of the carrier tone with respect to the \qam{N} signal. Note that at low \gls{CSPR}, the minimum-phase condition is occasionally violated, leading to reconstruction errors \cite{mecozzi_kramers_2016}. Between the \SI{1}{GBaud} 1\% roll-off \gls{RRC} \qam{N} signal with frequency components up to \SI{0.505}{GHz} and the digitally-inserted carrier tone at \SI{0.516}{GHz}, a small gap of only \SI{11}{MHz} is left. An investigation revealed this relatively small gap to balance the trade-off between increased \gls{KK} reconstruction errors due to increased \gls{SSBI} at smaller gaps versus increased \gls{KK} reconstruction errors due to bandwidth limitations of the employed receiver at larger gaps. With other equipment, the optimum may differ.

The \gls{MP} \qam{N} test signal is amplified and combined by a \gls{WSS} with a dummy band of 79 \SI{32}{GBaud} \gls{QPSK} signals, spaced according to the \SI{50}{GHz} ITU-T grid with a gap at \SI{1550.116}{nm} for the test signal to emulate full C-band transmission, see the schematic \textit{Spectrum} inset in \cref{fig:setup}. The \gls{WDM} signal is amplified and transmitted over the 100-span straight-line link. The average span length is \SI{100}{km} with \glspl{WSS} placed every 20 spans to flatten the transmission spectrum. Note that this is not a recirculating loop experiment, therefore, the link physically contained \SI{10000}{km} of fiber. The transmission fiber was designed for submarine transmission and has a loss of \SI{0.154}{dB \per km} and an effective mode-area of \SI{112}{\micro\meter\squared}. The total launch power at the input of each span is \SI{20}{dBm} and cannot be varied due to setup constraints. To investigate the influence of launch power on nonlinearities and noise, the launch power of the test signal relative to the dummy channels is varied using a \gls{VOA}. Monitoring taps at various points along the link allowed for evaluation at distances up to \SI{10000}{km}. At the receiver, the tapped \gls{WDM} signal is amplified and the test signal is filtered using a \SI{5}{GHz} optical bandpass filter. Another \gls{EDFA} amplifies the test signal with a \gls{VOA} controlling the power into a \SI{6.5}{GHz} photodiode. The electrical signal is sent to the workstation and digitized by a 12-bit \SI{4}{GS/s} \SI{1}{GHz} \gls{ADC} and processed in real-time on the \gls{GPU}, which has a maximum rated power of \SI{250}{W}.

\begin{figure}
\includegraphics[width=\linewidth]{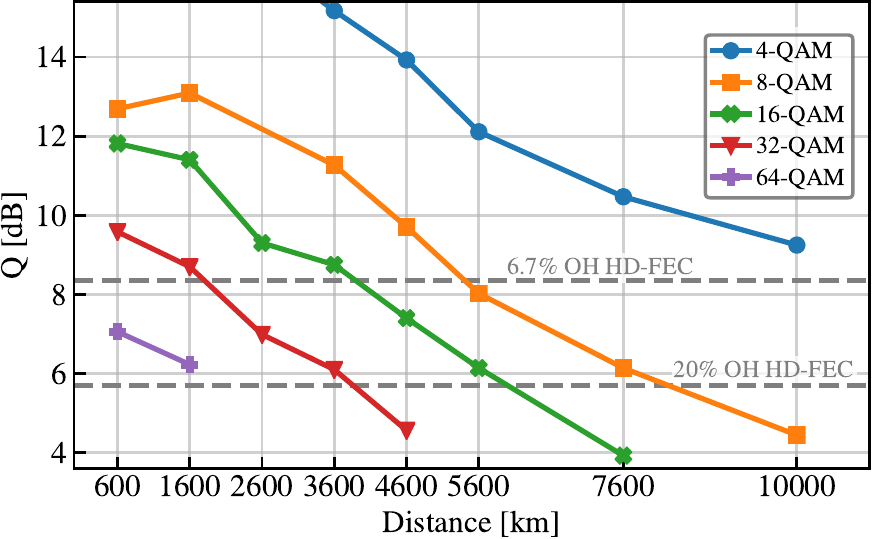}
\caption{Q-factor versus distance for various \qam{N} formats.}
\label{fig:distance} 
\end{figure}

The real-time \gls{DSP} is schematically shown in \cref{fig:setup} and its GPU-based implementation is described in detail in \cite{gpuJLT}. The \gls{ADC} transfers buffers of samples to the \gls{GPU} using \gls{DMA} over an 8-lane PCIe Gen3 interface. Each buffer contains 2\textsuperscript{22} samples and is subdivided into 8192 blocks for 100\% overlap-save \gls{FD} processing. Separate \gls{GPU} processing streams are used for each buffer to facilitate greater parallelization. The signal processing starts with a \gls{GPU} kernel handling the overlap with the last block of the previous buffer to make a contiguous stream of samples available for signal processing. This kernel also converts the samples from 12-bit fixed point to 32-bit floating point representation and handles the \gls{KK} front-end operations consisting of the square root and logarithm operations. Then, a 1024-point 100\% overlap-save \gls{FFT} pair enables a \gls{FD} Hilbert transformation to retrieve the phase of the optical signal. Note that these operations, through the utilization of millions of threads, leverage the massive parallelization offered by \glspl{GPU}, next to the parallelization offered by parallel processing streams. The retrieved phase is combined with the amplitude to reconstruct the optical field \cite{mecozzi_kramers_2016}, which is subsequently downshifted to DC for further processing. Another pair of \glspl{FFT} enables \gls{FD} static equalization and resampling from 4 to 2 samples-per-symbol. The static equalizer is optimized off-line using 203 taps at 2 samples per symbol. Finally, the signal is filtered by 4-tap adaptive \gls{DDLMS} widely-linear \cite{Silva_WidelyLinear} \gls{TD} equalizer capable of correcting for modulator IQ-imbalances. The decisions of the adaptive equalizer are demapped and transferred to \gls{RAM}. Since the \gls{GPU} is able to process incoming digitized buffers at a rate faster than they are produced by the \gls{ADC}, this receiver is considered operating in real time. Note that overlap between buffers is carefully handled as explained here and in \cite{gpuJLT}, enabling a fully contiguous data stream to be processed and transferred to \gls{RAM}, differentiating this receiver from a very fast block-wise off-line \gls{DSP}. Large buffers allow for many threads to process data in parallel and reduce overhead penalties, leading to lower development effort but increased latency. Future implementations can lower latency through careful optimization of buffer size and efforts to reduce overhead. Although \gls{GPU}-based \gls{FEC} decoding is possible \cite{Li_realtime_LDPC,suzuki_fec}, it is considered beyond the scope of this work. However, to limit \gls{FEC} complexity, we only assume hard decision \glspl{FEC}. Since the storage media in our workstation computer are not fast enough to save the data stream in real time, results were saved to \gls{RAM}. The size of our \gls{RAM} limited the real time trace lengths reported in this paper to about 21 seconds.
\section{Results and Discussion}
\label{sec:results}

\begin{figure}
\includegraphics[width=\linewidth]{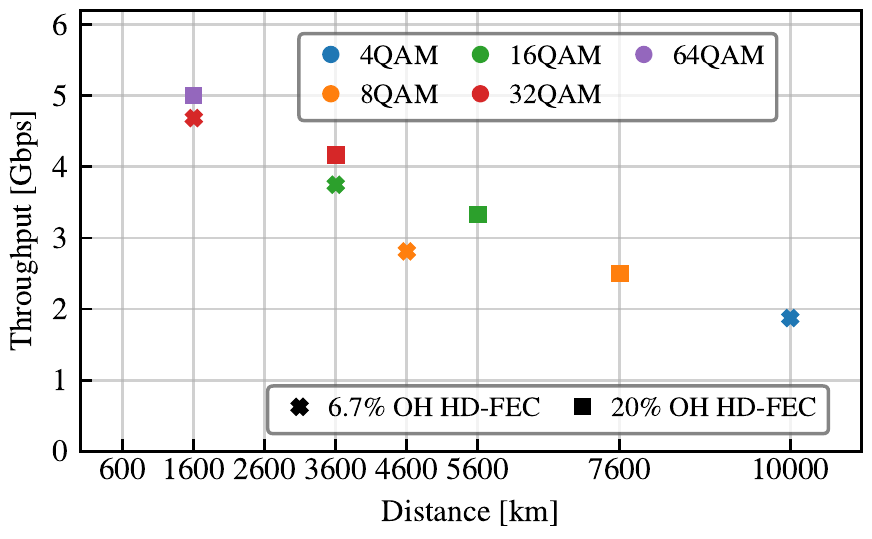}
\caption{Net throughput versus distance.}
\label{fig:throughputplot} 
\end{figure}

\cref{fig:distance} shows the Q-factor as a function of distance for all considered \qam{N} formats. \gls{CSPR} and relative launch power were optimized for each data point. \qam{4} reaches the 6.7\% \gls{OH} \gls{HDFEC} \cite{agrell_information-theoretic_2018} threshold at \SI{10000}{km} and since the physical link was not longer, we could not measure farther. If a 20\% \gls{OH} \gls{HDFEC} \cite{agrell_information-theoretic_2018} is employed, \qam{8} can be successfully transmitted over up to \SI{7600}{km}. Using this same threshold, \qam{16-, 32-, 64} can reach \SIlist{5600;3600;1600}{km}, respectively. The net throughput for all formats and both \gls{HDFEC} thresholds is plotted against distance in \cref{fig:throughputplot}. Note that this is not interpolated, the farthest data point above the \gls{HDFEC} threshold is plotted for each format. Using the software-defined flexible multi-modulation format receiver, a suitable format can be chosen for each distance, maximizing the net throughtput. 

\begin{figure}
\includegraphics[width=\linewidth]{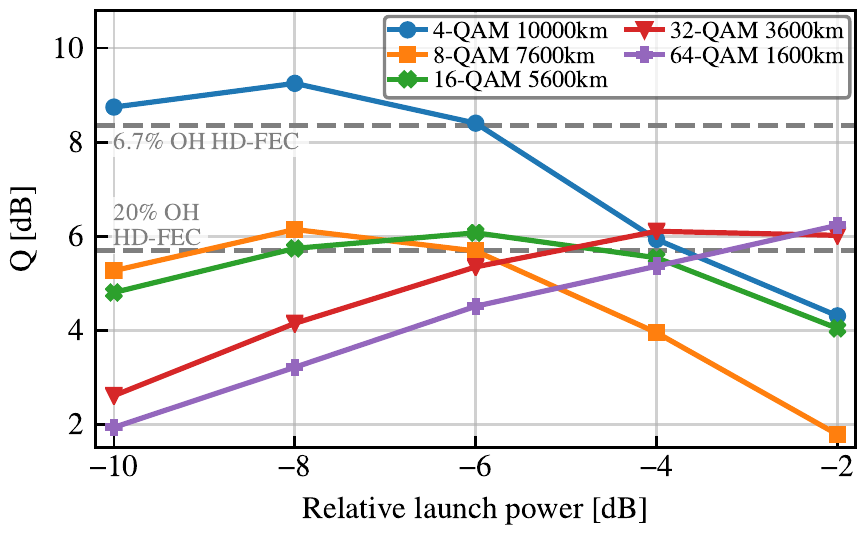}
\caption{Q-factor versus relative launch power for various formats at their FEC-limit distance.}
\label{fig:lpvsq} 
\end{figure}

\begin{figure}
\includegraphics[width=\linewidth]{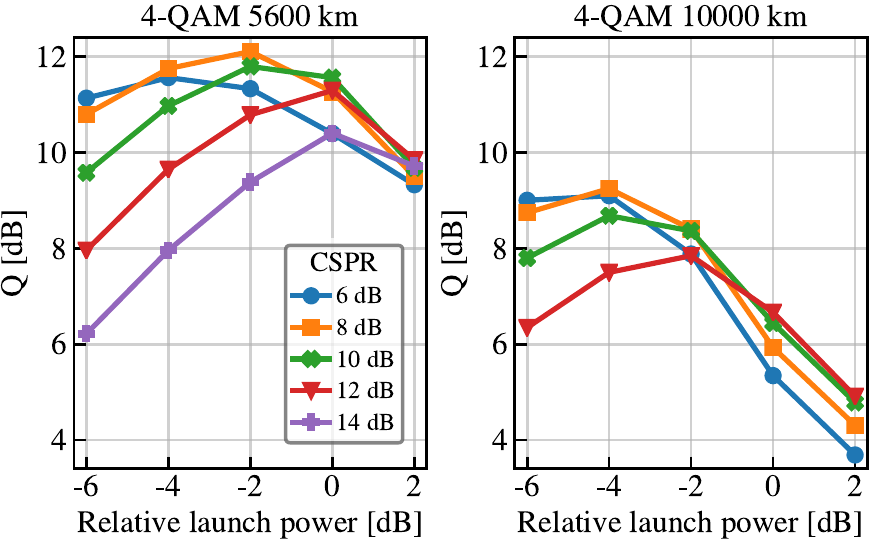}
\caption{Q-factor versus relative launch power for \qam{4}.}
\label{fig:qpsklaunchpower} 
\end{figure}

\cref{fig:lpvsq} shows the Q-factor versus relative launch power for all considered modulation formats. It shows that the power of the test channels needs to be well below that of the dummy channels for maximum performance. This suggests a dominant impact of \gls{SPM}, perhaps on the \gls{MP} carrier tone. The optimum shifts with format and distance. Therefore, this effect is investigated for \qam{4} at \SIlist{5600;10000}{km} in \cref{fig:qpsklaunchpower} for various \glspl{CSPR}. At \SI{10000}{km}, the optimal relative launch power for \qam{4} is about \SI{2}{dB} lower than at \SI{5600}{km}. This confirms the optimal launch power is dependent on distance. 

\cref{fig:qpsklaunchpower} also shows the \gls{CSPR} trade-off between nonlinearities and noise. Since a higher \gls{CSPR} has more power in the carrier tone and less power in the \qam{N} signal, the influence of noise is increased. Therefore, the optimal launch power is higher for higher \glspl{CSPR} as shown in the figure. At high launch powers, nonlinear interactions degrade the signal leading to a drop in performance. 

The optimum launch power dependence on transmission distance is further investigated in Fig. 6, where the relative launch power with optimized CSPR leading to the highest Q-factor is plotted as a function of distance for both \qam{4-, and 32}. For both formats, the optimal launch power decreases substantially with transmission distance. This is a surprising outcome since in conventional coherent transmission systems, optimal launch power tends to remain independent of transmission distance because signal degradation due to fiber nonlinearities and noise scale at a similar rate for a fully populated C-band \cite{GNmodel}. However, in the case of an \gls{MP} \qam{N} signal detected using a \gls{KK} coherent receiver, the nonlinear interactions between signal and carrier tone seem to scale at a higher rate than noise with distance, leading to a lower optimal launch power with greater transmission distance. Further investigation into \gls{MP} signal degradation due to nonlinear fiber interactions is required to fully describe this behavior.

\begin{figure}
\includegraphics[width=\linewidth]{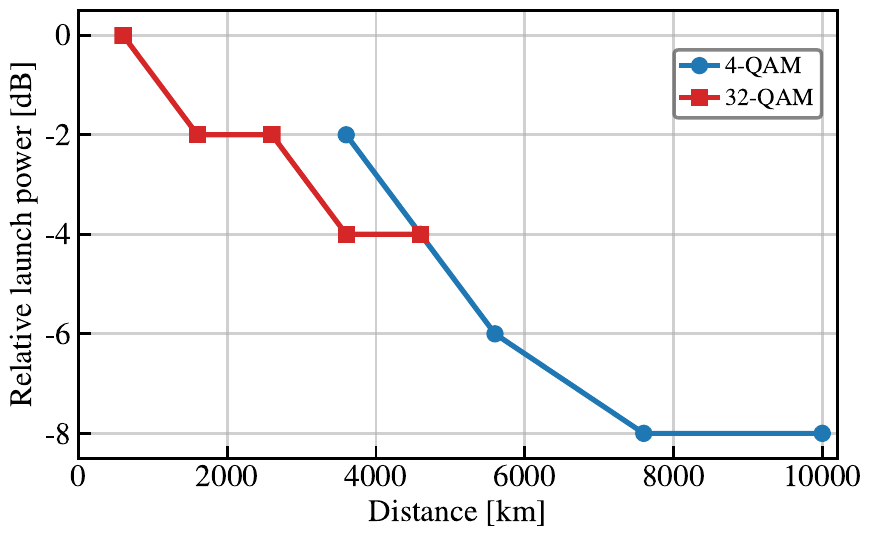}
\caption{Optimal relative launch power versus distance.}
\label{fig:optimallaunchpower} 
\end{figure}

\begin{figure}
\includegraphics[width=\linewidth]{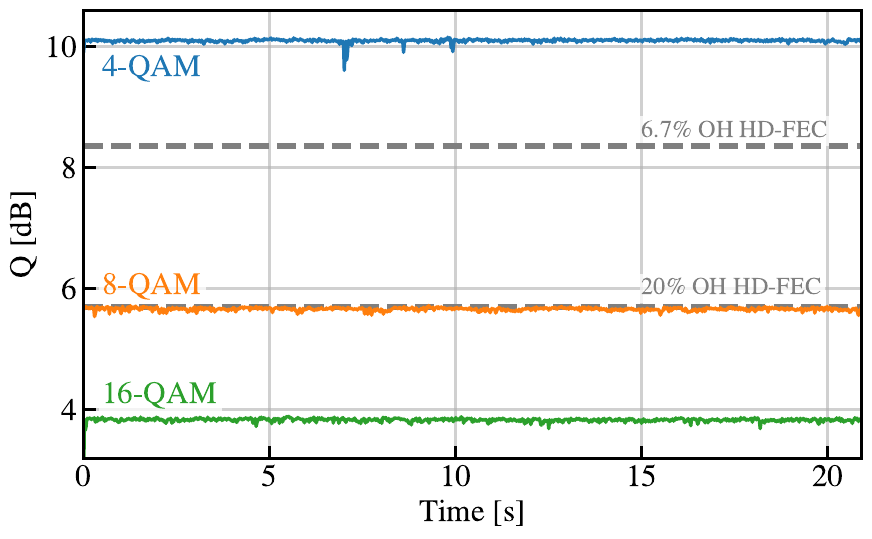}
\caption{Continuous real-time traces at \SI{7600}{km}.}
\label{fig:longtraces} 
\end{figure}

Finally, \cref{fig:longtraces} shows continous real-time traces for \qam{4-, 8-, and 16} at \SI{7600}{km} transmission distance. The Q-factor is estimated from \gls{BER} and is averaged over intervals of \SI{21}{ms}. The 21 second traces show stable short-term averaged Q-factors, even for \qam{16} which operates in a very low Q-factor regime, demonstrating the robustness of the signal processing. This shows the receiver is capable of sustained operation with the main limitation imposed by the \gls{RAM} capacity of the workstation used for the experimental validation. Power consumption of the receiver concept is expected to decrease as energy efficiency of \glspl{GPU} continues to improve.

\section{Conclusions}
\label{sec:conclusions}

A flexible real-time multi-modulation format receiver using a commercial off-the-shelf \gls{GPU} is experimentally evaluated for long-haul optical communication. \SI{1}{GBaud} \glsfirst{MP} \qam{N} signals are successfully transmitted over an experimental \SI{10000}{km} straight-line link and detected using a \gls{KK} coherent receiver. The real time signal processing includes the \gls{KK} algorithm as well as digital chromatic dispersion compensation, enabling successful transmission of \qam{4} up to \SI{10000}{km}. \qam{8-, 16-, 32-, 64} reach \SIlist{7600;5600;3600;1600}{km} of transmission, respectively. \Gls{CSPR} and launch power are optimized for each format and transmission distance. The result of this optimization is analyzed and we show that the nonlinear behavior of \gls{MP} signals is substantially different from conventional coherent systems. Furthermore, we show that our system is capable of sustained operation in 21 second long continuous real-time traces, showing stable performance of \qam{4-, 8-, and 16} over \SI{7600}{km}. These results demonstrate the potential for employing \glspl{GPU} in long-haul optical transmission systems.

\ifCLASSOPTIONcaptionsoff
  \newpage
\fi



%


\printbibliography[notcategory=ignore]

%








\end{document}